\documentclass[aps,prl,twocolumn,superscriptaddress,showpacs]{revtex4-1}
\usepackage{graphicx,amsmath,amssymb,bm}
\usepackage[utf8]{inputenc}

\newcommand{\heff}{H_{\rm eff}}
\newcommand{\veff}{V_{\rm eff}}
\newcommand{\fmi}{\, \text{fm}^{-1}}
\newcommand{\mev}{\, \text{MeV}}
\newcommand{\kev}{\, \text{keV}}
\newcommand{\gevi}{\, \text{GeV}^{-1}}
\newcommand{\hw}{\hbar\omega}
\newcommand{\df}{d_{5/2}}
\newcommand{\dt}{d_{3/2}}
\newcommand{\so}{s_{1/2}}

\newcommand{\stn}{S_{\rm 2n}}
\newcommand{\stp}{S_{\rm 2p}}

\begin{document}

\title{Exploring $\bm{sd}$-shell nuclei from two- and three-nucleon 
interactions \\ with realistic saturation properties}

\author{J.\ Simonis}
\email[E-mail:~]{simonis@theorie.ikp.physik.tu-darmstadt.de}
\affiliation{Institut f\"ur Kernphysik, Technische Universit\"at
Darmstadt, 64289 Darmstadt, Germany}
\affiliation{ExtreMe Matter Institute EMMI, GSI Helmholtzzentrum f\"ur
Schwerionenforschung GmbH, 64291 Darmstadt, Germany}

\author{K.\ Hebeler}
\email[E-mail:~]{kai.hebeler@physik.tu-darmstadt.de}
\affiliation{Institut f\"ur Kernphysik, Technische Universit\"at
Darmstadt, 64289 Darmstadt, Germany}
\affiliation{ExtreMe Matter Institute EMMI, GSI Helmholtzzentrum f\"ur
Schwerionenforschung GmbH, 64291 Darmstadt, Germany}

\author{J.\ D.\ Holt}
\email[E-mail:~]{jholt@triumf.ca}
\affiliation{TRIUMF, 4004 Wesbrook Mall, Vancouver, British Columbia, V6T 
2A3 Canada}

\author{J.\ Men\'{e}ndez}
\email[E-mail:~]{menendez@nt.phys.s.u-tokyo.ac.jp}
\affiliation{Department of Physics, University of Tokyo, Hongo, Tokyo 113-0033, Japan}

\author{A.\ Schwenk}
\email[E-mail:~]{schwenk@physik.tu-darmstadt.de}
\affiliation{Institut f\"ur Kernphysik, Technische Universit\"at
Darmstadt, 64289 Darmstadt, Germany}
\affiliation{ExtreMe Matter Institute EMMI, GSI Helmholtzzentrum f\"ur
Schwerionenforschung GmbH, 64291 Darmstadt, Germany}

\begin{abstract}
We study ground- and excited-state properties of all $sd$-shell nuclei
with neutron and proton numbers $8 \leqslant N,Z \leqslant 20$, based on a set
of low-resolution two- and three-nucleon interactions that predict
realistic saturation properties of nuclear matter. We focus on
estimating the theoretical uncertainties due to variation of the
resolution scale, the low-energy couplings, as well as from the
many-body method. The experimental two-neutron and two-proton
separation energies are reasonably well reproduced, with an
uncertainty range of $\sim 5 \mev$. The first excited $2^+$ energies
also show overall agreement, with a more narrow uncertainty range
of $\sim 500 \kev$. In most cases, this range is dominated by
the uncertainties in the Hamiltonian.
\end{abstract}


\maketitle

\paragraph{Introduction.} 

Recent advances in nuclear theory have established the importance of
three-nucleon (3N) forces in understanding the structure of
medium-mass nuclei, for the evolution to the neutron and proton
driplines~\cite{Otsu10Ox,Hage12Ox3N,Holt13PR,Cipo13Ox,Herg13MR} and
the formation of shell
structure~\cite{Holt12Ca,Hage12Ca3N,Gall12Ca,Holt13Pair,Wien13Nat,Soma14GGF2N3N,Bind14CCheavy,Herg14MR}.
Three-nucleon forces are also key for realistic saturation properties
of nuclear
matter~\cite{Bogn05nuclmat,Hebe11fits,Hage13ccnm,Carb13nm,Cora14nmat},
which in turn are obtained from global analyses of all nuclei. To date,
ab initio studies of medium-mass nuclei have largely focused on
closed-shell nuclei or isotopic chains, generally in the vicinity of
semi-magic nuclei, and no comprehensive study exists to explore
nuclear forces over a full range of the nuclear chart, such as the
$sd$ shell.

An additional challenge is the quantification of theoretical
uncertainties~\cite{Doba14error}. Calculations in oxygen and calcium
isotopes based on nuclear forces derived from chiral effective field
theory (EFT)~\cite{Epel09RMP,Mach11PR} suggest that the uncertainties
from the many-body methods are well
controlled~\cite{Herg13MR,Cipo13Ox,Herg14MR,Soma14GGF2N3N,Hage14rev}. Therefore,
uncertainties in the input Hamiltonian, such as truncations in the
chiral EFT expansion or uncertainties in the low-energy couplings,
likely remain the dominant source of uncertainty. Note that recently,
first studies of the statistical uncertainties from numerically optimized chiral
forces \cite{Ekst13optNN,Ekst15sat} and to quantify correlations between
chiral EFT couplings \cite{Ekst15unc,Carl15sim} have been performed.

In this work we investigate all 
$sd$-shell nuclei based on chiral two-nucleon (NN) and 3N interactions 
with realistic saturation properties. We derive microscopic 
valence-space Hamiltonians, which we diagonalize to obtain 
ground-state energies, two-neutron and two-proton 
separation energies, and first excited  $2^+_1$ energies. By varying 
the resolution scale in nuclear forces and the low-energy 3N couplings, 
we provide theoretical uncertainty estimates for these observables.
In addition we also explore the uncertainty associated due to the many-body
calculations. We find that the resulting energy ranges, which 
are in good agreement with experimental data throughout the region, are
mainly driven by uncertainties in the Hamiltonian.

\paragraph{Nuclear interactions.}

At the NN level, we start from the next-to-next-to-next-to-leading
order (N$^3$LO) $500 \mev$ potential of Entem and Machleidt
(EM)~\cite{Ente03EMN3LO}. We then use the similarity renormalization
group (SRG)~\cite{Bogn07SRG,Bogn09PPNP} to evolve this interaction to
a series of low-resolution scales $\lambda_{\rm NN} = 1.8, 2.0, 2.2
\fmi$. Taking chiral EFT as a general low-momentum basis, we combine
each SRG-evolved NN interaction with the leading N$^2$LO 3N
forces~\cite{Kolc94fewbody,Epel02fewbody}, where the $c_i$ couplings
in the two-pion-exchange 3N interaction are taken consistently with
the NN interaction: $c_1 = -0.81 \gevi$, $c_3 = -3.2 \gevi$, $c_4 =
5.4\gevi$. We also vary independently the 3N cutoff $\Lambda_{\rm 3N}
= 2.0, 2.5 \fmi$ for the $\lambda_{\rm NN} = 2.0 \fmi$ interaction. In
addition, to probe uncertainties in the $c_i$ couplings, we use 3N
forces with the $c_i$ values obtained from the Nijmegen NN partial
wave analysis (PWA): $c_1 = -0.76 \gevi$, $c_3 =-4.78 \gevi$, $c_4 =
3.96 \gevi$~\cite{Rent03ciPWA} for the $\lambda_{\rm NN}/
\Lambda_{\rm 3N} = 2.0/2.0 \fmi$ interaction. For all Hamiltonians,
the low-energy couplings $c_D$, $c_E$ in the 3N one-pion-exchange and
3N contact interaction have been fit to the $^3$H binding energy and
$^4$He charge radius using Faddeev- and Fadeev-Yakubowsky
calculations~\cite{Hebe11fits}. With this set of five NN+3N
interactions, which predict nuclear saturation properties within
uncertainties~\cite{Hebe11fits}, we explore all $sd$-shell nuclei. The
different interactions are denoted as $1.8/2.0$~(EM), $2.0/2.0$~(EM),
$2.2/2.0$~(EM), $2.0/2.5$~(EM), $2.0/2.0 \fmi$~(EM+PWA), where the
labeling indicates $\lambda_{\rm NN}/\Lambda_{\rm 3N}$ and the $c_i$
couplings used.

\paragraph{Microscopic valence-space Hamiltonians.}

Based on the above interactions, we construct an effective
valence-space Hamiltonian, $\heff=\sum_i \varepsilon_i a^{\dagger}_i
a_i + V_{\rm eff}$, where $\varepsilon_i$ denote the single-particle
energies (SPEs), and $\veff$ is the effective two-body interaction for
valence nucleons.  With $\heff$ we perform valence shell-model
calculations, where the many-body problem is solved exactly for the
particles in the valence space on top of a closed core.

Recently, nonperturbative methods for calculating $\heff$ in
medium-mass nuclei have been
developed~\cite{Tsuk12SM,Bogn14SM,Jans14SM,Dikm15NCSMSM} and applied
to the oxygen and very recently fluorine~\cite{Cace1524F} isotopes. In
this work, we use many-body perturbation theory
(MBPT)~\cite{Hjor95MBPT}, which provides a diagrammatic order-by-order
expansion for SPEs and $\veff$, to take into account many-body
processes outside the valence space. At third order, MBPT based on
RG/SRG-evolved interactions with low cutoffs shows a reasonable
order-by-order convergence for SPEs and $\veff$ in medium-mass
nuclei~\cite{Holt14Ca}. To explore uncertainties associated with MBPT,
we study valence-shell Hamiltonians obtained at second- and
third-order MBPT. In contrast to phenomenological
interactions, such as USDA/B~\cite{Brow06USD} that fit both SPEs and $\veff$
to experimental $sd$-shell data, our results are without adjustments.
Therefore, we do not expect to reach accuracies comparable to the best fit USD interactions.

Studies of oxygen~\cite{Holt13Ox,Caes1326O} and
calcium~\cite{Gall12Ca,Wien13Nat,Holt14Ca} isotopes showed that
extending the valence space beyond one major shell provides additional
binding and can lead to improvements for neutron-rich systems.
However since our main interest is to perform a comprehensive study of
$sd$-shell nuclei, estimating the theoretical uncertainty associated
with the initial Hamiltonians, we limit our valence space to the $sd$
shell ($\df$, $\dt$, and $\so$ proton and neutron single-particle
orbitals on top of a $^{16}$O core). We work in a harmonic-oscillator
basis with $\hw=13.53 \mev$, appropriate for the $sd$ shell, and scale
all matrix elements of $\veff$ and bound SPEs by $A^{-1/3}$ to correct
for the increase in nuclear size. For all $\lambda_{\rm NN}$ considered, the
calculations are converged in a basis consisting of 13 major shells
for NN forces~\cite{Holt14Ca}.  For 3N forces, we allow a total energy
of the three single-particle states up to 12$\hbar\omega$
($E_{3\,\text{max}}=12$) in a basis of 13 major shells.

\begin{figure}
\begin{center}
\includegraphics[width=0.9\columnwidth,clip=]{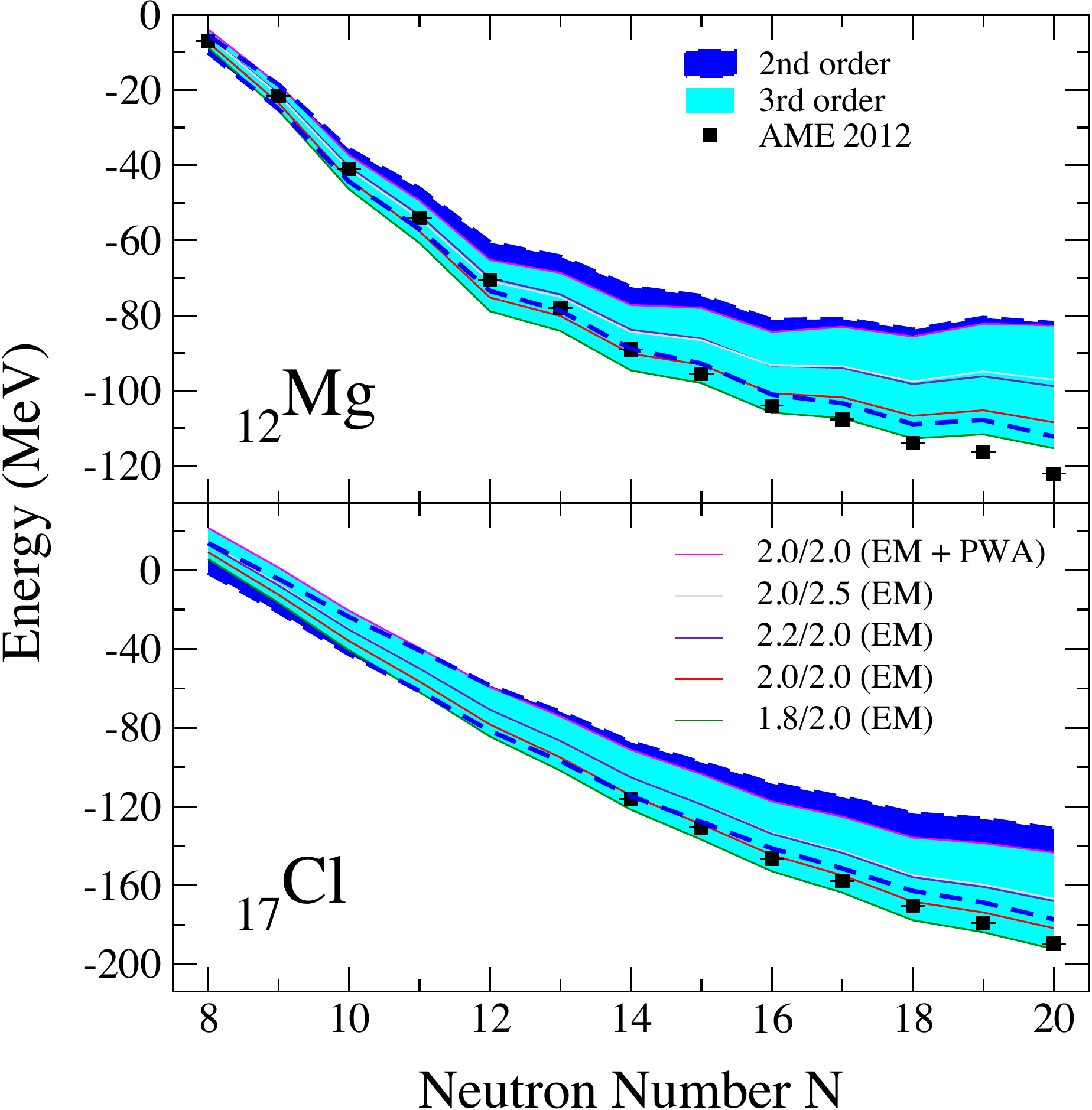}
\end{center}
\caption{Ground-state energies of the magnesium (top) and chlorine 
isotopes (bottom panel) relative to $^{16}$O at second (blue, darker
band) and third order (cyan, lighter band) in MBPT and compared to the
Atomic Mass Evaluation (AME 2012)~\cite{Wang12AME12}. The uncertainty
bands are spanned by the five different NN+3N interactions (see text
for details). The ordering in the legend is with decreasing
ground-state energies.\label{gs}}
\end{figure}

\begin{figure*}
\begin{center}
\includegraphics[width=0.9\textwidth,clip]{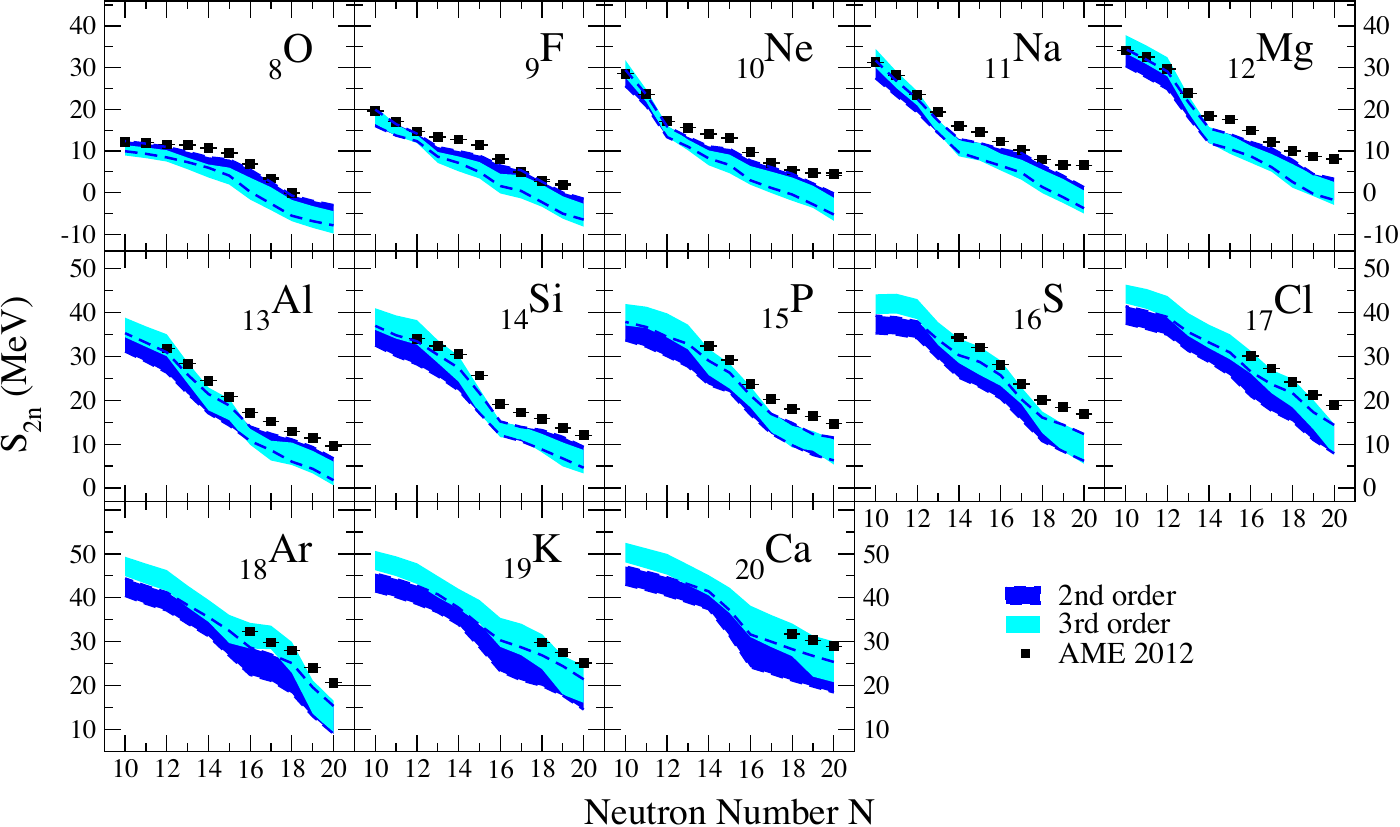}
\end{center}
\caption{Uncertainty estimates for the two-neutron separation energies 
$\stn$ of $sd$-shell isotopic chains at second (blue, darker band) and 
third order (cyan, lighter band) in MBPT and compared to the Atomic 
Mass Evaluation (AME 2012)~\cite{Wang12AME12}.\label{S2n}}
\end{figure*}

\paragraph{Results.}

Figure~\ref{gs} shows the ground-state energies of magnesium and
chlorine isotopes compared to the Atomic Mass Evaluation (AME
2012)~\cite{Wang12AME12}. The second- and third-order MBPT results are
represented by the blue, darker and the cyan, lighter bands,
respectively, where the width of each band is spanned by the five
different NN+3N interactions considered.

The experimental ground-state energies for magnesium isotopes are
generally within our uncertainty band, with neutron-rich isotopes at
the lower side. Only the most neutron-rich isotopes are underbound in
our calculations. On the other hand, the ground-state energies of all
chlorine isotopes are in good agreement with our uncertainty band, and
they are typically within the lower side of the the third-order MBPT
band, defined by the $\lambda_{\rm NN}/\Lambda_{\rm 3N} = 1.8/2.0$,
$2.0/2.0$, $2.2/2.0$, $2.0/2.5 \fmi$ interactions.  In general we find
better agreement between our results and experiment for the isotopic
chains of heavier elements, which suggests a somewhat too weak
neutron-neutron interaction in our $sd$-shell calculations. This was
also observed in Refs.~\cite{Holt13Ox,Caes1326O} for the oxygen
isotopes.

The estimated uncertainties in calculated ground-state energies are
dominated by the different input Hamiltonians. Specifically, the
resolution-scale dependence by varying $\lambda_{\rm NN}$ from $1.8 -
2.2 \fmi$ with $\Lambda_{\rm 3N}=2.0\fmi$ is somewhat larger than the
$\Lambda_{\rm 3N}$ dependence from $2.0 - 2.5 \fmi$ for $\lambda_{\rm
NN}=2.0\fmi$. This results in a combined resolution-scale dependence of
approximately $1.0 \mev$ per valence particle in $^{32}$Mg and
$^{37}$Cl. When also including the PWA $c_i$ values in 3N forces, the
uncertainty roughly doubles to about $2.0 \mev$ per valence particle.
For both second- and third-order MBPT bands, the $2.0/2.0 \fmi$ (EM+PWA) interactions
generally define the least bound calculations (for ground-state
energies, the upper end of the bands).

The difference between second- and third-order MBPT results is
relatively small compared to the width of each band, indicating a
reasonable, but still incomplete, convergence of the MBPT approach in
this region. For magnesium and chlorine, third-order results are more
bound because of more attractive proton-neutron interactions, whereas
for oxygen (not shown), second-order results are more bound than at
third order mainly due to the neutron single-particle energies.
When the uncertainty associated to the MBPT is also included, the
total uncertainty increases to $2.1 \mev$ and $2.8 \mev$ per valence
particle in $^{32}$Mg and $^{37}$Cl, respectively.

Figure~\ref{S2n} compares theoretical and experimental two-neutron
separation energies $\stn$ for all isotopic chains from oxygen to
calcium ($Z=8-20$). The theoretical calculations describe the overall
experimental trends reasonably well, but in general our uncertainty
bands underestimate the empirical values.  This is especially the case
in lighter elements and for the most neutron-rich nuclei for all
isotopic chains. This is probably related to the underbinding of the
$sd$-shell calculations when valence neutron-neutron interactions are
dominant. We also note that around $N=20$, the ground states of
$^{29,30}$Ne~\cite{Bell05ne28,Yana03ne30},
$^{30,31}$Na~\cite{Trip07na30,Door10na31}, and
$^{31,32}$Mg~\cite{Seid11mg31,Moto95mg32be2} are dominated by deformed
configurations not captured in our $sd$-shell calculations (this is
the so-called island of inversion). Consequently, our bands do not
reproduce the change in slope of $\stn$ around $N=20$ for Ne, Na, or Mg.

Similar to the ground-state energies, the dominant uncertainties arise
from the different Hamiltonians, with smaller differences between
second- and third-order MBPT results. Typical the uncertainty range
for $\stn$ is $\sim 5 \mev$. The exception are $N<Z$ isotopes, more
visible in heavier elements, where the difference between second- and
third-order results is comparable to the uncertainty between input
Hamiltonians, due to too weak proton-neutron interactions at
second-order MBPT, adding up to a total uncertainty of $\sim 10 \mev$.

\begin{figure*}
\begin{center}
\includegraphics[width=0.9\textwidth,clip=]{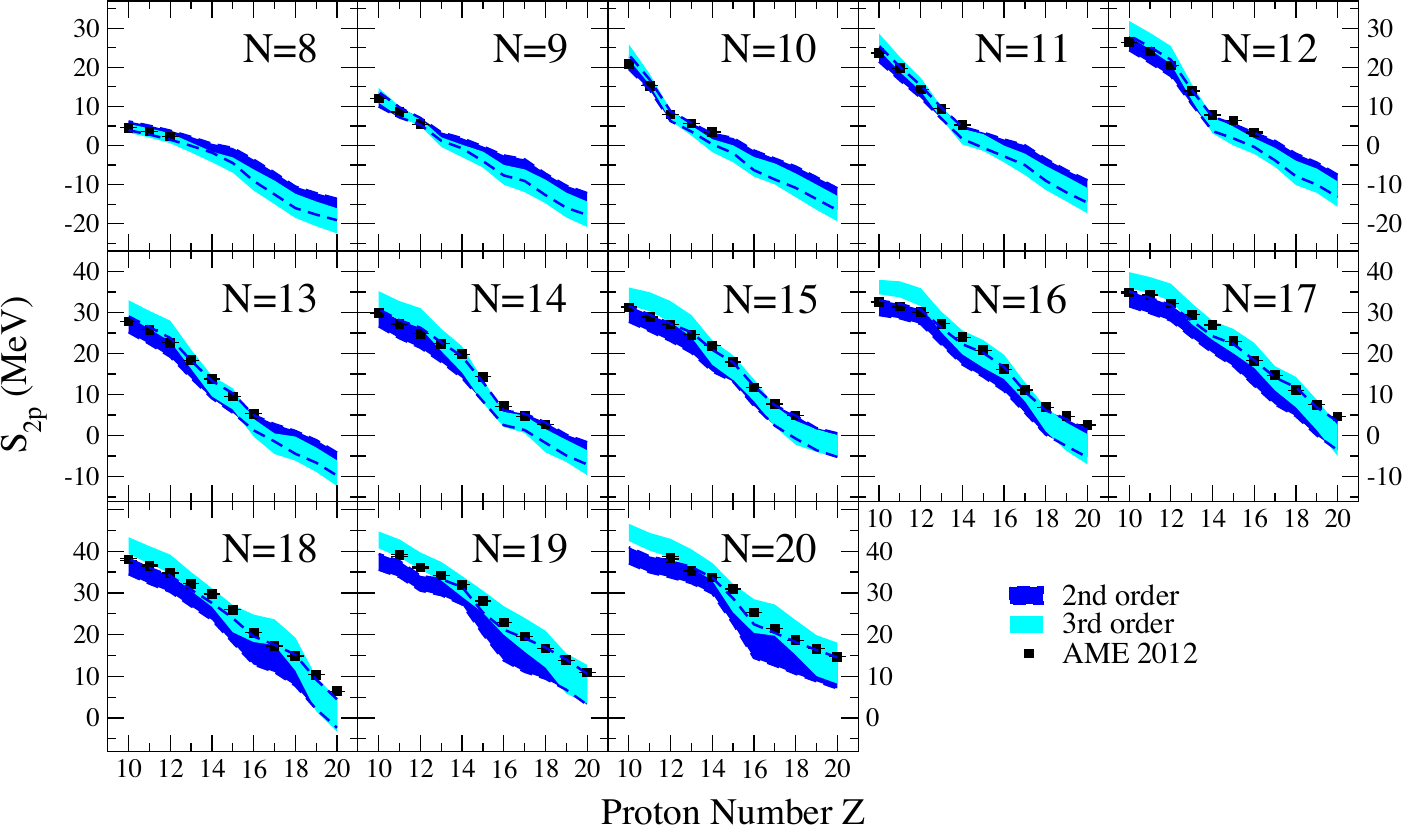}
\end{center}
\caption{Uncertainty estimates for two-proton separation energies $\stp$
of $sd$-shell isotonic chains at second (blue, darker band) and third
order (cyan, lighter band) in MBPT and compared to the Atomic Mass
Evaluation (AME 2012)~\cite{Wang12AME12}.\label{S2p}}
\end{figure*}

In Fig.~\ref{S2p} we show the two-proton separation energy $\stp$ for
all isotonic chains from $N=8$ to $N=20$. Our results agree very well
with experiment in all cases, and remarkably most experimental values
fall within the third-order MBPT band. Only in few proton-deficient
and very proton-rich nuclei do experimental $\stp$ lie within the
second-order band.  Since there are fewer proton-rich nuclei known
experimentally than neutron-rich nuclei, $\stp$ are in general
informative about proton-neutron interactions.  The much better
agreement in $\stp$ than for $\stn$ compared to experiment suggest
that the different Hamiltonians considered capture better (mostly
isoscalar) proton-neutron interactions than neutron-neutron
interactions.  Again, the sensitivity to the input Hamiltonians
dominates the theoretical $\stp$ uncertainties (with a similar range
of $\sim 5 \mev$), except for proton-deficient nuclei where the MBPT
uncertainty is comparable (with a total uncertainty of $\sim 10 \mev$).

\begin{figure*}
\begin{center}
\includegraphics[width=0.8\textwidth,clip=]{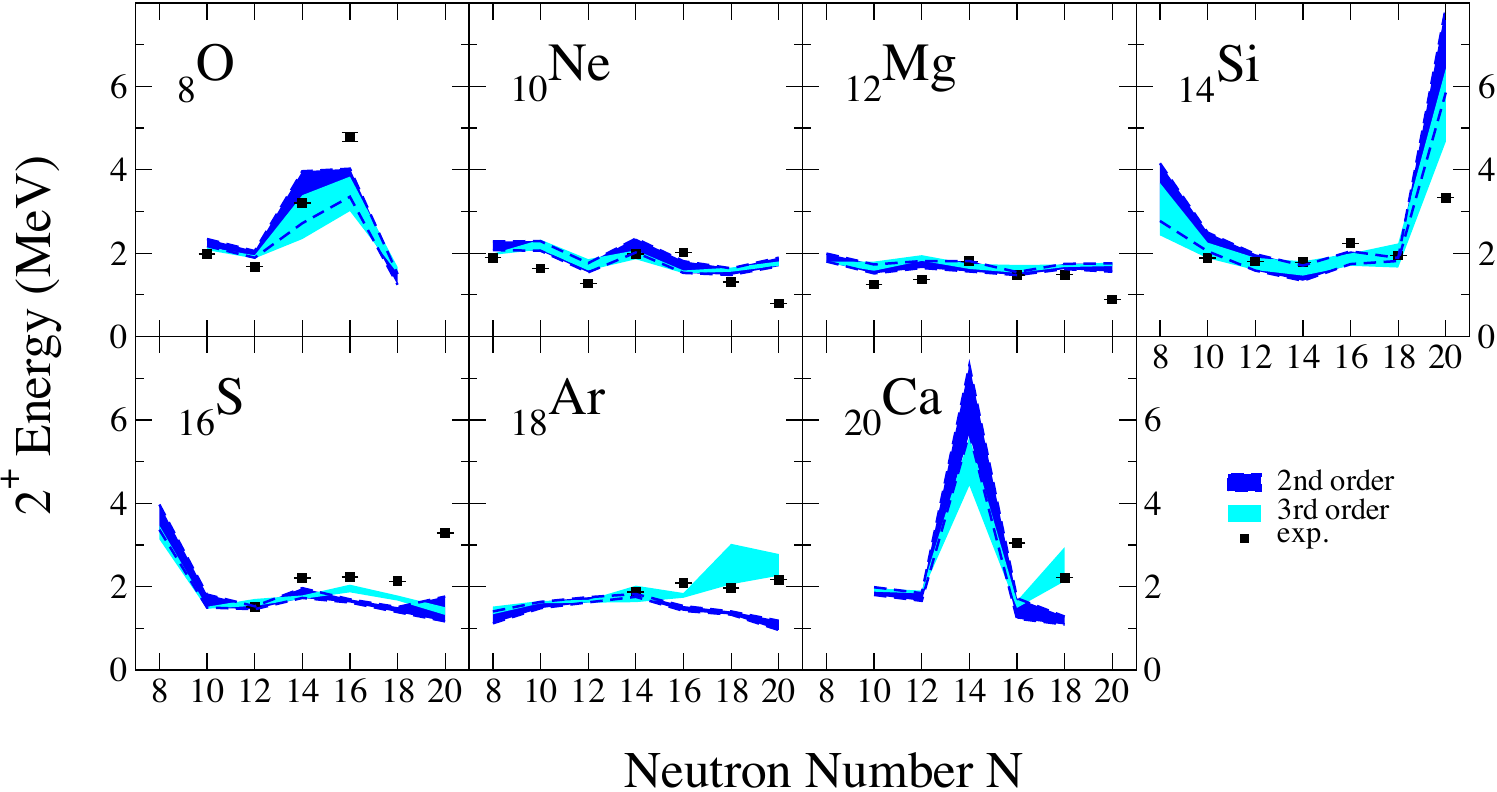}
\end{center}
\caption{Uncertainty estimates for excitation energies of the first
2$^+_1$ states in even-even $sd$-shell isotopes at second (blue, darker 
band) and third order (cyan, lighter band) and compared to experimental
data from ENSDF~\cite{nndc14ENSDF}.\label{2+}}
\end{figure*}

Finally in Fig.~\ref{2+}, the calculated first excited $2^+_1$
energies are compared to experimental data for all even-even
$sd$-shell isotopes.  The spread of the uncertainty band is typically
smaller than $\sim 500\kev$, with generally reasonable agreement to
experiment. However, in the cases with high-lying $2^+_1$ states,
indicative of shell closures ($^{22}$O, $^{24}$O, $^{22}$Si,
$^{34}$Si, $^{34}$Ca), the uncertainty can be as large as $\sim 1
\mev$. This means that, while our bands in general predict shell
closures consistently, the actual excitation of the $2^+_1$ state is
very sensitive to the details of the input Hamiltonian. The width of
the uncertainty band is mostly due to the $2.0/2.0 \fmi$ (EM+PWA)
interaction, which is also responsible for the unusually large
uncertainty band in $^{36}$Ar. In general, the second- and third-order
MBPT bands mostly overlap, except for $N \sim Z$ argon and calcium
isotopes, where only third-order MBPT results are in agreement to
experiment. Similar to the case of $\stn$, we also note that the
$2^+_1$ states within the island of inversion, the $N=20$ isotopes
$^{30}$Ne~\cite{Yana03ne30} and $^{32}$Mg~\cite{Moto95mg32be2}, are
deformed, and their relatively low excitation energies cannot be well
described in our $sd$-shell calculations.

\paragraph{Summary.}

We have presented a comprehensive study of ground-state energies,
$\stn$, $\stp$, and first excited 2$^+_1$ energies for all $sd$-shell
nuclei: isotopic chains from oxygen to calcium and isotonic chains
from $N=8$ to $N=20$. This is based on NN+3N Hamiltonians that have
been fitted only to $A=3,4$ nuclei that predict realistic saturation
properties of nuclear matter, without additional adjustments.  We have
focused on estimating the theoretical uncertainties due to the
different input Hamiltonians and associated with the many-body
calculations.  We find reasonable agreement to experimental data,
especially in nuclei dominated by valence proton-neutron interactions.
For neutron-rich systems, calculations in extended valence spaces are
needed, due to too weak neutron-neutron interactions in the $sd$ shell.

Generally, we find that the estimated theoretical uncertainties are
dominated by differences in the NN+3N interactions.  While the present
first study is limited to the EM 500 MeV N$^3$LO NN potential, it will
be important to perform more comprehensive studies.  These efforts
will incorporate the exploration of different fitting procedures for
the 3N low-energy constants $c_D, c_E$, different regulator forms and
cutoff values for NN and 3N interactions as well as order-by-order
convergence studies in the chiral EFT
expansion~\cite{Epel15improved,Epel14NNn4lo,Hebe15n3lopw,Furn15uncert}.  For
improved studies of the resolution scale dependence, we plan to
perform calculations based on consistently SRG-evolved NN and 3N
interactions~\cite{Jurg09SRG3N,Hebe12msSRG}.  Improving our
uncertainty estimates due the many-body calculation is more
challenging, because corrections to MBPT (i.e., fourth-order
contributions) are not attainable at present.  However,
nonperturbative methods for valence-space Hamiltonians can provide a
controlled framework to assess the many-body
approximations~\cite{Tsuk12SM,Bogn14SM,Jans14SM,Dikm15NCSMSM}.

\begin{acknowledgments}
We thank H.~Hergert, N.~Tsunoda, and T.~Otsuka for very useful
discussions. This work was supported by the ERC Grant No. 307986
STRONGINT, the BMBF under Contract No. 05P15RDFN1, the National
Research Council of Canada and NSERC, by an International Research
Fellowship from the Japan Society for the Promotion of Science (JSPS),
and JSPS KAKENHI grant No.~$26\cdot04323$.  Computations were
performed with an allocation of computing resources at the J\"ulich
Supercomputing Center.
\end{acknowledgments}

\bibliography{sdshell_uncertainties}

\end{document}